\documentclass[a4paper,11pt]{article}
\pdfoutput=1 

\usepackage{jheppub} 
                     
\usepackage{braket,slashed,bm}

\usepackage{simplewick}
\usepackage[T1]{fontenc} 

\def\nn{\nonumber\\ }
\def\rd{{\rm d}}
\def\abs#1{\left| #1 \right| }

\def\lchi{\Lambda_\chi}
\def\lqcd{ \Lambda_{\text{QCD}}}
\def\vev#1{\left\langle #1 \right\rangle}
\def\msbar{\overline{\hbox{MS}}}

\def\bea{\begin{eqnarray}}
\def\eea{\end{eqnarray}}
\def\ba{\begin{eqnarray}}
\def\ea{\end{eqnarray}}
\def\be{\begin{equation}}
\def\ee{\end{equation}}

\title{
On Gauge Invariance and Minimal Coupling
}

\author[a]{Elizabeth E.~Jenkins,}

\author[a]{Aneesh V.~Manohar,}

\author[b,1]{Michael Trott}\note{Corresponding author.}

\affiliation[a]{Department of Physics, University of California at San Diego, 9500 Gilman Drive,\\ La Jolla, CA 92093-0319, USA}
\affiliation[b]{Theory Division, Physics Department, CERN, CH-1211 Geneva 23, Switzerland}

\emailAdd{ejenkins@ucsd.edu}
\emailAdd{amanohar@ucsd.edu}
\emailAdd{michael.trott@cern.ch }

\abstract{The principle of minimal coupling has been used in the study of Higgs boson interactions to argue that certain higher dimensional operators in the low-energy effective theory generalization of the Standard Model are suppressed by loop factors, and thus smaller than  others. It also has been extensively used to analyze beyond-the-Standard-Model theories.
We show that in field theory, and even in quantum mechanics,  the concept of minimal coupling is ill-defined and inapplicable as a general principle, and give many pedagogical examples which illustrate this fact.  We also clarify some related misconceptions about the dynamics of strongly coupled gauge theories. Many arguments in the literature on Higgs boson interactions that use minimal coupling, particularly in pseudo-Goldstone Higgs theories, are inherently flawed.}

\begin{document} 
\maketitle

\section{Introduction}
Given the discovery of a Higgs-like boson at the LHC with a mass of $125$ GeV, it is important
to have a model-independent effective field theory (EFT) approach to Higgs boson physics.  To date, the Standard Model (SM) provides a good description of LHC data up to energies of around 1~TeV.  No evidence of beyond the SM (BSM)  physics, natural or otherwise, has emerged thus far, so it is important not to introduce strong bias when considering how BSM physics can impact experimental studies, and to keep the analysis as general as possible.

The properties of the recently discovered scalar are consistent with it being identified with the Higgs boson of the SM, but these properties have not been measured with precision yet.  In this paper, we will assume that the discovered scalar boson can be described by the SM plus BSM effects parametrized by adding 
higher-dimensional operators in the SM fields to the SM Lagrangian.  The higher-dimensional operators are suppressed by powers of a 
high-energy scale $\Lambda$.  The leading operators which affect the Higgs production and decay amplitudes arise at dimension six, and so are suppressed by $1/\Lambda^2$.  Since no BSM states have been found so far, LHC results already indicate that the scale $\Lambda$ is higher than the scale $v= 246\,\text{GeV}$ of electroweak symmetry breaking (EWSB). 

A recent general classification of higher dimensional operators involving SM fields is given in 
Ref.~\cite{Grzadkowski:2010es}, which showed that there are 59 independent dimension-six operators (assuming the conservation of baryon number).   Ref.~\cite{Grzadkowski:2010es} reduces the set of possible dimension-six operators from those of the earlier work 
Ref.~\cite{Buchmuller:1985jz} by using the classical equations of motion to eliminate redundant operators.  While the choice of operator basis is not unique, the operators of Ref.~\cite{Grzadkowski:2010es} represent one consistent choice of independent operators.
In our recent paper~\cite{Grojean:2013kd}, we studied a subset of these dimension-six operators
which modify the $h \to \gamma \gamma$ and $h \to Z \gamma $ decay rates, and  calculated the renormalization group evolution of these operators, including operator mixing.\footnote{The running and mixing of dimension-six operators modifying the SM has been extensively studied in the literature, see e.g.\  
Ref.~\cite{Hagiwara:1993ck,Hagiwara:1993qt,Alam:1997nk,Arzt:1992wz,Weinberg:1989dx,Wilczek:1979hc,Braaten:1990gq,Degrassi:2005zd,Dekens:2013zca} for previous studies.}  Our calculation shows that it is important to include operator mixing of the new physics operators for precision analysis of Higgs decays.  Deviations of $h \rightarrow \gamma \gamma$ from the SM rate at the $\sim 10\%$ level are possible due to renormalization group running of new physics operators contributing to the decay.

These results were criticized recently by Ref.~\cite{Elias-Miro:2013gya},
which argued that (i) the specific choice of operator basis is crucial for the analysis, and there is a special basis which makes the analysis very simple, (ii) there is a powerful and general classification of EFT coefficients based on a  ``tree'' and ``loop''  classification, even when the underlying theory is strongly interacting and non-perturbative, and (iii) the principle of minimal coupling\footnote{Minimal coupling is the replacement of an ordinary derivative $\partial_\mu$ by the covariant derivative $D_\mu=\partial_\mu + i g A_\mu$ to construct a theory with gauge interactions from a theory without gauge interactions.}
defines an unambiguous classification scheme in which higher-dimensional operators which violate minimal coupling are suppressed by loop  factors of $g^2/(16\pi^2)$, where $g$ is a coupling constant.\footnote{Ref.~\cite{Grojean:2013kd} used this classification for the sole purpose of an illustrative example (based on the strongly interacting light Higgs model~\cite{Giudice:2007fh}) of mixing effects.}  If true, these arguments have significant implications for the construction of EFTs, including widely studied EFTs such as chiral perturbation theory ($\chi$PT), heavy quark effective theory (HQET), the operator analysis of weak decays, etc.\    

Recent use of minimal coupling stems from the analysis of Ref.~\cite{Giudice:2007fh}, where minimal coupling is asserted to underly a ``tree'' and ``loop'' operator classification scheme for higher-dimension Higgs boson interactions with other SM fields.  
The analysis of Ref.~\cite{Giudice:2007fh} has been quite influential, and it has led to significant phenomenological work on BSM Higgs boson interactions~\cite{Giudice:2007fh,Low:2009di,Elias-Miro:2013gya,Falkowski:2013dza,Contino:2013kra}.
It also underlies the arguments in Ref.~\cite{Elias-Miro:2013gya}.

Since the idea of EFT was pioneered by Weinberg and others in the 1960s and 1970s however, minimal coupling has not generally been used as an organizational scheme for higher dimensional operators, and it has long been appreciated that minimal coupling is of limited 
value in the context of EFTs~\cite{weinberglectures}.  Even earlier, H.~Weyl appreciated that theories could have non-minimally coupled interactions~\cite{Weyl:1929fk,PhysRev.77.699} in his study of electrodynamics and gravity. In fact, such interactions were required in these references for mathematical consistency.\footnote{We thank S.~Deser for this reference.}

In this paper, we show that the principle of minimal coupling is ill-defined in field theory and even in quantum mechanics, and that in general there is no unambiguous classification of higher dimensional operators into tree and loop operators based on a principle of minimal coupling.  We provide many examples of well-studied EFTs to illustrate these points.
Furthermore, we explain why it is not possible to postulate that an EFT satisfies a minimal coupling principle by the choice of a suitable UV theory.  This point is relevant when considering the generation of higher dimensional operators by a UV theory, as many UV theories considered in the literature are themselves non-renormalizable EFTs with no valid operator classification based on a principle of minimal coupling. Many of the ideas in this paper can be traced at least as far back as 1970~\cite{weinberglectures} to Weinberg's distinction between algebraic and dynamical symmetries.
While many of the ideas are familiar to EFT practitioners, we hope that at least a few points are novel.

The outline of the paper is as follows:
\begin{itemize}

\item In Sec.~\ref{sec:EFTpowercounting}, we review the standard terminology of EFT, the notion of tree and loop graphs in the EFT, and EFT power counting. In some recent work invoking minimal coupling operator classification schemes, terms such as ``tree'', ``loop'', and ``power counting'' do not have the traditional definitions used in the EFT literature, so we summarize the standard EFT nomenclature.

\item In Sec.~\ref{sec:tree}, we show that it is not possible to classify amplitudes in a low-energy effective theory arising from a strongly coupled gauge theory in a ``tree'' and ``loop'' classification based on the diagrams involving the fundamental fields. We also comment on the weakly coupled case, and show that the classification depends on assumptions about the high energy theory.

\item 
In Sec.~\ref{sec:minimal}, we show how the
usual definition of minimal coupling is ambiguous, and does not lead to a clear-cut distinction between operators which preserve or violate minimal coupling. We propose several natural generalizations, which suffer from the same problem. 

\item In Sec.~\ref{minimal}, we provide many field-theoretic and quantum mechanical examples which violate the minimal coupling classification.

\item Sec.~\ref{sec:chpt} discusses the concept of minimal coupling in chiral perturbation theory and in pseudo-Goldstone Higgs  theories.

\item Other common misconceptions are examined in Sec.~\ref{sec:other}, i.e.:
\begin{itemize}

\item strongly coupled gauge theories can  be treated as weakly coupled theories in which coupling constants are taken to be of order $4\pi$, 

\item particles must have magnetic moments with $g=2$ (at tree-level) in an EFT for good high-energy behavior up to the high-energy scale $\Lambda$ where the EFT  breaks down.

\item one can adjust UV theories so that their IR effective theories are minimally coupled.

\end{itemize}

We explain why these notions are not true in Sec.~\ref{sec:other}.

\item Finally, in the appendix, we show that any dimension-six Higgs operator can be generated by tree-level exchange, if the underlying theory is an EFT.

\end{itemize}

\section{EFT and Power Counting}\label{sec:EFTpowercounting}

We start by summarizing well-known properties of EFTs (for a review, see Ref.~\cite{Manohar:1996cq}). This summary will introduce the notation and terminology used in this paper. An EFT is described by a local gauge-invariant Lagrangian written as an expansion in operators,
\begin{align}
\mathcal{L} &= \mathcal{L}_{d \le 4} + \frac{\mathcal{L}_5}{\Lambda} +  \frac{\mathcal{L}_6}{\Lambda^2} + \ldots
\end{align}
where $\mathcal{L}_{d \le 4}$ denotes terms of dimension less than or equal to four, $\mathcal{L}_5$ consists of terms of dimension five, etc.\
In certain cases, there also can be topological terms, which contribute to the action, but are not given as the integral of a Lagrange density.

An EFT can be used to compute particle interactions in a power series in $1/\Lambda$. The $1/\Lambda$ amplitude is given by one insertion of $\mathcal{L}_5$, the $1/\Lambda^2$ amplitude by one insertion of $\mathcal{L}_6$ or two insertions of $\mathcal{L}_5$, etc.\
Dimensionful parameters in the EFT consist of particle masses and momenta, so that an amplitude in the EFT has an expansion in powers of 
$p/\Lambda$, which is referred to as the EFT power counting parameter.  One can also formulate a power counting expansion for fields and operators.
The EFT is weakly coupled at $p \rightarrow 0$.  Tree and loop graphs in the EFT are well-defined, but this designation does not correspond to tree and loop graphs in the underlying theory.  Whenever we refer to tree and loop graphs, we will mean tree and loop graphs computed using the EFT Lagrangian in the weak coupling regime of the EFT.

Loop graphs in the EFT can be divergent, and are usually renormalized using the $\msbar$ scheme, in which operators of different dimension do not mix and no positive powers of $\Lambda$ are generated by loop graphs.  The EFT power counting is \emph{explicit}, i.e.\ one can read off the powers of $1/\Lambda$ from the operator insertions.  The renormalization structure of the EFT follows from the power counting.  Let $\{c_{4}\}$, $\{c_5\}$, etc.\ be the coupling constants in $ \mathcal{L}_{4} $, $\mathcal{L}_5$, etc, then schematically,
\begin{align}\label{2.2}
\mu \frac{ \rd \{c_5\}}{\rd \mu} &= A_1\left(\{c_{4}\}\right)\ \{c_5\}, \nn
\mu \frac{ \rd \{c_6\}}{\rd \mu} &= A_2\left(\{c_{4}\}\right) \{c_6\}  + A_3 \left(\{c_{4}\}\right) \ \{c_5\}^2,
\end{align}
and so on. This pattern shows that operators of arbitrarily high dimension in the EFT are required to absorb divergences from loops with arbitrarily many insertions of $\mathcal{L}_{d \ge 5}$.  For this reason, EFTs are referred to as non-renormalizable theories. The infinite set of higher dimension operators is not generated if there are no terms with dimension $\ge 5$;  these are the renormalizable theories, and are a very special case of EFT with $\Lambda \to \infty$.  The pattern of Eq.~(\ref{2.2}) shows that, in general, 
it is not useful to introduce an ordering scheme of the $c_i$ of the EFT based on some external principle  
unless the ordering follows from a symmetry, and hence is preserved by loop diagrams.  Minimal coupling, as we discuss below, is not a symmetry, and it does not lead to an ordering scheme which is preserved under renormalization.

It is important to remember that the EFT has a different divergence structure from the full theory, because massive propagators are expanded out in a momentum expansion,
\begin{align}
\frac{1}{p^2-M^2} &= - \frac{1}{M^2} + \frac{p^2}{M^4} + \ldots
\end{align}
The EFT is constructed by a procedure known as matching: coefficients in the EFT are adjusted order by order in the power counting expansion so that the EFT produces the same $S$ matrix elements as the full theory.  Operators in the full theory need to be matched onto operators in the effective theory.  

Fields in the effective theory are not identical to the corresponding fields in the full theory.  For example, the gluon field $A_\mu^a$ in QCD with six quark flavors is not the same as the gluon field $A_\mu^{\prime a}$ in the EFT of QCD with five quark flavors obtained when the top quark is integrated out of the full QCD theory.  As an example, the matching result for the field strength tensor is~\cite{Chetyrkin:1997un}
\begin{align}
{G_{\mu \nu}^AG^{A\mu \nu} \over2 g} &= {G_{\mu \nu}^{\prime A} G^{\prime A\mu \nu}  \over2 g^\prime} {\partial g^\prime \over \partial g}
-\sum_i {\partial m_i^\prime \over \partial g} \bar \psi_i^\prime \psi_i^\prime,
\end{align}
where the unprimed fields are in the full QCD theory with a top quark, and the primed fields are in the EFT without a top quark.  The EFT should be thought of as a \emph{different} field theory than the original full theory; it has been constructed to reproduce the $S$-matrix elements of the original theory order by order in the power counting expansion in $1/\Lambda$.
The input of the underlying theory is via the matching procedure. In some cases, a  perturbative matching can be performed, as in HQET.

By construction, higher dimension operators in the EFT are treated as insertions, i.e.\ the effective theory is treated as an expansion in $1/\Lambda$. The expansion parameter $p/\Lambda$ becomes order unity, and the EFT expansion breaks down, when $p$ becomes order 
$\Lambda$.
While the EFT is treated as an expansion in $1/\Lambda$, the theory need not be perturbative. For example, the standard analysis of hadronic weak decays treats the weak interactions using four-fermion operators with a coefficient $G_F \sim 1/v^2$. The weak Hamiltonian can be treated perturbatively, but the hadronic matrix elements of the weak Hamiltonian needed for hadronic weak decays are non-perturbative, and one expands the decay amplitude in $G_F \lqcd^2$.

The EFT describes the interactions of low-energy degrees of freedom, such as  particle scattering amplitudes at momenta below $\Lambda$. In general, an EFT is constructed by writing down the most general possible Lagrangian consistent with the symmetries and the low-energy degrees of freedom, with arbitrary coefficients for the operators. One can estimate the size of the coefficients using naive dimensional analysis~\cite{Manohar:1983md}, or relate coefficients through symmetries, but their numerical values generally have to be determined from experiment if the UV theory is strongly coupled, \emph{or if the UV theory is not known.}

The above power counting procedure works for massless or massive gauge theories, as long as the gauge boson mass is generated by spontaneous symmetry breaking~\cite{Georgi:1985kw}.  Thus, one can use an EFT for the SM augmented by higher dimension operators suppressed by $\Lambda > v $, and this EFT will be valid up to energies of order $\Lambda$, since the EFT includes the Higgs field which gives mass to the electroweak gauge bosons by spontaneous symmetry breaking.

\section{Tree versus Loop }\label{sec:tree}

Weakly coupled theories have a perturbative expansion in coupling constants, and amplitudes can be expanded in the number of loops.  Typically, each additional loop gives a $1/(16\pi^2)$ suppression, in addition to suppressions from factors of small coupling constants, which are generically denoted by $g^2$ in this paper.  The leading amplitudes are tree amplitudes, followed by one-loop amplitudes, two-loop amplitudes, etc.  This weak coupling analysis does not carry over to strongly coupled gauge theories, such as QCD at low energies.  
To illustrate this point, consider QCD with light quark masses set to zero. The theory is given by a coupling constant $g_3(\mu)$, which evolves with $\mu$. As is well-known, QCD has the property of dimensional transmutation --- $g_3(\mu)$ can be replaced by a dimensionful parameter $\lqcd$. In terms of the one-loop $\beta$-function,
\begin{align}
\mu \frac{\rd g_3}{\rd \mu} &=  -\frac{b_0 g_3^3}{16\pi^2}, 
\end{align}
$\lqcd$ and $g_3(\mu)$ are related by
\begin{align}
\left( \frac{\lqcd}{\mu} \right)^{b_0} &= e^{-8\pi^2/\left[ \hbar\, g_3^2(\mu)\right]}\, ,
\label{lqcd}
\end{align}
where we have put back the factors of $\hbar$~\cite{Coleman:1975qj,Coleman:1978ae}. $\lqcd$ is a non-perturbative parameter.
It is clear from Eq.~(\ref{lqcd}) that
$\lqcd$ is not of any given order in the loop expansion. It cannot even be thought of as summing the $g_3^2$ expansion to all orders.
There is an essential singularity at $g_3^2\hbar=0$ in the $g_3^2$ and $\hbar$ expansions, and $\lqcd$ is non-analytic in $g_3^2 \hbar$. A formal expansion of Eq.~(\ref{lqcd}) in powers of $g_3^2>0$ gives a series with all terms vanishing. It simply is not possible to describe non-perturbative physics as  a series of factors of $g^2/16 \pi^2$ with $g \rightarrow 4 \pi$, and treat QCD as a weakly coupled theory in which $g \sim 4 \pi$.

The low-energy dynamics of QCD is governed by $\chi$PT, which is discussed further in Sec~\ref{sec:chiral}. The scattering amplitudes depend on the pion decay constant $f_\pi \propto \lqcd$, and they cannot be characterized as $n$-loop amplitudes for some value of $n\ge0$.  $\chi$PT has its own power counting as an EFT, but this expansion in $p/\Lambda_\chi$ has nothing to do with the loop expansion of QCD.  
Loops in the effective theory defined by the chiral Lagrangian are not related to loops in the full theory, QCD. (The only connection between the two theories is through the flavor dependence of amplitudes in the large $N_c$ limit~\cite{Manohar:1998xv} in terms of single vs multi-trace operators, but that is not what is being referred to here.)

In weakly coupled theories, one can classify operators based on whether they are generated by tree or loop graphs when heavy particles are integrated out.  The distinction between tree and loop operators depends on the high energy theory. For example, in the Standard Model, nonleptonic weak decays of hadrons can be described by four-quark operators.  $\Delta S=1$ weak decays such as $K \to \pi \pi$ are given by the current-current operator
\begin{align}
\mathcal{L}_{\Delta S=1}&= - \frac{4 G_F}{\sqrt 2} V_{ud} V_{us}^*\ \overline u \gamma^\mu P_L d\ \overline s \gamma_\mu P_L u\,,
\label{3.3}
\end{align}
which is generated at tree level by single $W$ exchange at the scale $M_W$.
$\Delta S=2$ processes, such as $K^0-\overline K^0$ mixing, are given by very similar-looking current-current operators
\begin{align}
\overline s \gamma^\mu P_L d\ \overline s \gamma_\mu P_L d\,.
\end{align}
However, in the Standard Model, these operators are not generated at tree level because there are no flavor changing neutral currents at tree-level due to the GIM mechanism. They are generated by box graphs, and are second order in the weak interactions.
 They could be generated at tree-level if there were new physics interactions such as new gauge bosons or quark multiplets which violated the GIM mechanism. The tree-loop classification of operators in the EFT requires knowledge of the underlying high energy theory, even in a weak coupled theory.

The $\Delta S=1$ operator Eq.~(\ref{3.3}) looks like the product of currents. However, it is a composite operator in an interacting field theory, and cannot be treated as the product of two current operators when one takes hadronic matrix elements. A very prominent feature of nonleptonic weak decays is the  $\Delta I=1/2$ rule: the $\Delta I=1/2$ amplitude $A_{1/2}$ is enhanced relative to the $\Delta I=3/2$ amplitude $A_{3/2}$ by a factor of $A_{1/2}/A_{3/2} \sim 20$. If the $\Delta S=1$ operator is treated as the product of currents, one can show that $A_{1/2}=\sqrt 2 A_{3/2}$ (see Prob.~4.1 in~\cite{Manohar:1998xv}).

In summary, when non-perturbative physics is present (and sometimes even when it is not), intuition based on weak coupling or minimal coupling, can fail in an EFT in non-intuitive ways. 
These cautionary remarks are relevant to the case of the pseudo-Goldstone boson (PGB) Higgs theories, and to other theories containing  higher dimensional operators.

\section{Minimal Coupling is Ambiguous}\label{sec:minimal}

Usually, in constructing effective theories, one writes down all possible gauge invariant operators up to a given dimension.  All   operators occurring at a given dimension are regarded as equally important because they contain the same suppression factor
of $1/\Lambda^{d-4}$. 
Recently, it has become popular to advocate an additional ordering principle, minimal coupling.  One assumes that the underlying UV theory is minimally coupled.  Some gauge invariant operators result from integrating out particles in the minimally coupled theory at tree level, while others do not, and the latter operators have coefficients suppressed by at least a loop suppression factor of $g^2/(16\pi^2)$ beyond the usual EFT power counting.  For this concept to have any content, it is necessary that some gauge invariant operators are minimally coupled, and others are not. Otherwise, minimal coupling becomes indistinguishable from gauge invariance.

To be able to use minimal coupling in EFTs, it is important to define minimal coupling at the operator level in the EFT in an unambiguous way. Otherwise, different authors using the same principle will obtain different results.  Unfortunately, the definition of minimal coupling in EFT is ambiguous, as explained below.

The principle of minimal coupling is a method of  constructing theories with gauge interactions from a theory without gauge interactions.
In non-relativistic quantum mechanics, one constructs the Hamiltonian with electromagnetic interactions from the Hamiltonian without electromagnetic interactions by the replacement
\begin{align}
\mathbf{p} \to \bm{\pi} = \mathbf{p}- e q \mathbf{A},
\label{min}
\end{align}
where $\mathbf{A}$ is the photon field, and $e$ is the coupling constant. The momentum operator $\mathbf{p}$ acts on particle fields (or wavefunctions), and $q$ is the corresponding particle charge. In four-vector notation, one replaces the ordinary derivative by the covariant derivative,
\begin{align}
i \partial_\mu \to iD_\mu = i \partial_\mu - e q A_\mu.
\label{min2}
\end{align}
In non-Abelian gauge theories, $i \partial_\mu$ is replaced by the covariant derivative $iD_\mu = i \partial_\mu - g T^a A_\mu^a$, where the matrices $T^a$ are the generators of the gauge group in the representation of the field on which the covariant derivative acts. 
A covariant derivative $D_\mu$ is only defined when it acts on some field, because the  matrices $T^a$ in $D_\mu$ depend on the representation of the field.

It is important to keep in mind that minimal coupling is merely one way of constructing a gauge Lagrangian from an ungauged Lagrangian, and that the minimal coupling prescription does not always lead to a unique gauge Lagrangian, or even the correct one.
Elevating ``minimal coupling'' to a fundamental principle and confusing it with gauge invariance is a mistake. One can always go the other way --- the ungauged theory is obtained from the gauged theory by setting all the gauge fields and gauge couplings to zero. The question is whether a mapping in the opposite direction can be defined: can a unique
gauge theory Lagrangian be obtained from a Lagrangian without gauge interactions? The answer to this question is ``no,'' as we show below.

The difference between gauge invariance and minimal coupling can be illustrated by a simple example.  In QED, consider the three 
gauge invariant dimension-six operators
\begin{align}
O_1 &= \phi^\dagger \phi D_\mu \phi^\dagger D_\mu \phi,  & O_2 &= e^2 \phi^\dagger \phi F_{\mu\nu}F^{\mu \nu},  & O_3 &= e^2\Phi^\dagger \Phi F_{\mu\nu}F^{\mu \nu},
\label{4.3}
\end{align}
where $\phi$ is a charged scalar field with $q=1$, and $\Phi$ is a neutral complex scalar field with $q=0$. The procedure for going from the gauge theory to the theory without gauge fields is well-defined --- set the gauge fields to zero. The ungauged operator analogues of the above three operators are
\begin{align}
\widetilde O_1 &= \phi^\dagger \phi \,\partial_\mu \phi^\dagger \partial_\mu \phi,  & \widetilde O_2 &= 0,  &\widetilde O_3 &= 0\,.
\label{4.4}
\end{align}
If we apply a principle of minimal coupling to Eq.~(\ref{4.4}), we recover $O_1$ from $\widetilde O_1$, but $O_{2,3}$ cannot be recovered from $\widetilde O_{2,3}$, which are zero.
In the standard EFT power counting, $O_{1,2,3}$ all have coefficients of order $1/\Lambda^2$.
Invoking the principle of minimal coupling at an operator level, on the other hand, implies that $O_{2,3}$ are suppressed by an additional loop suppression factor of $g^2/(16\pi^2)$ relative to $O_1$.

The above procedure for constructing a gauge invariant theory from a theory without gauge fields via minimal coupling is not unique, however.
The operator $\widetilde O_2$, while zero, can be written in the equivalent form
\begin{align}
\widetilde O_2 &= - \phi^\dagger  \left[\partial_\mu,\partial_\nu\right] \left[\partial^\mu,\partial^\nu \right] \phi\,.
\label{4.5}
\end{align}
Now, $\left[\partial^\mu,\partial^\nu\right]\phi=0$, but $\left[D^\mu,D^\nu\right]\phi=ie qF^{\mu \nu}\phi$, so $O_2$ can be recovered from 
$\widetilde O_2$
by a minimal coupling procedure, since $\phi$ has $q=1$.   
It is a logical possibility that the correct rule for obtaining a gauge theory from an ungauged theory is to write down all possible interactions in the non-gauge theory, including terms such as Eq.~(\ref{4.5}) which are zero, before applying a principle of minimal coupling.  This new rule gives a \emph{different} result from the previous paragraph --- now the coefficient of $O_2$ is not loop suppressed.  $O_3$ is still loop suppressed, because $\Phi$ is a \emph{neutral} field with $q=0$, so $D^\mu=\partial^\mu$ when it acts on $\Phi$.  No commutator gymnastics can produce an $F_{\mu \nu}$. Thus, there are two distinct ways to apply a principle of minimal coupling to the same theory, and these two possibilities give \emph{different} tree and loop classifications of higher dimensional operators.  To distinguish these two distinct ways of applying minimal coupling, we refer to the one in this paragraph as the principle of next-to-minimal coupling.

It is also instructive to study the case of a non-Abelian gauge theory.  Consider $SU(N)$ gauge theory with a scalar field $\varphi$ which transforms according to the fundamental representation.  There are two dimension-six gauge invariant operators analogous to $O_{2,3}$ of the previous QED case, namely
\begin{align}
O_4 &=\varphi^\dagger G_{\mu \nu} G^{\mu \nu} \varphi, & O_5 &= \varphi^\dagger \varphi\ \text{Tr}\, G_{\mu \nu} G^{\mu \nu},
\end{align}
where $G_{\mu\nu} \equiv G_{\mu \nu}^a T^a$.  These operators are not generated by the principle of minimal coupling, since they both vanish when the gauge field is set to zero.  The first operator can be generated by the principle of next-to-minimal coupling applied to the vanishing ungauged operator
\begin{align}
\widetilde O_4 &=  \left(\left[\partial_\mu , \partial_\nu\right]\varphi \right)^\dagger \left(\left[\partial_\mu , \partial_\nu\right]\varphi \right) \ .
\end{align}
The second operator $O_5$, however, cannot be generated by next-to-minimal coupling, because of the color structure.
The representation matrix $T^a$ in the covariant derivative $D_\mu = \partial_\mu + i g T^a A^a_\mu$ depends on the field on which it acts.
An object such as $\text{Tr}\,\left[D_\mu , D_\nu\right]\left[D_\mu , D_\nu\right]$ is meaningless, since the covariant derivatives need to act on some field on the right.  One can try constructing a $\widetilde O_5$ by acting on $\varphi^\dagger \varphi$, but $\varphi^\dagger \varphi$ is a gauge singlet, and $D_\mu$ acting on a gauge singlet is an ordinary derivative.  Acting on the adjoint $\varphi^\dagger T^a \varphi$ also does not work, because $D^\mu$ does not change the color transformation property of the object on which it acts, and a Lagrangian term must be invariant under color.  Once again, the principle of minimal coupling and the principle of next-to-minimal coupling give \emph{different} tree and loop classifications of operators.  The principle of minimal coupling implies that both $O_4$ and $O_5$ are loop suppressed, whereas the principle of next-to-minimal coupling implies that only $O_5$ is loop suppressed.

In summary, we have attempted to apply a principle of minimal coupling systematically in an EFT at an operator level.  The minimal coupling procedure is not unique, and different procedures give different tree versus loop operator classifications.

\section{EFTs are not Minimally Coupled}\label{minimal}

In constructing effective Lagrangians, one can impose symmetries such as baryon or lepton number conservation, or flavor symmetries.  These symmetries lead to relations between scattering amplitudes that must be respected by the effective theory, and hence lead to constraints on the effective theory.  This statement is true even if the flavor symmetry is spontaneously broken, as in $\chi$PT.  
Weinberg~\cite{weinberglectures} refers to such symmetries as algebraic symmetries because they lead to algebraic relations between $S$-matrix elements.  In contrast, Weinberg~\cite{weinberglectures} calls gauge invariance  a dynamical symmetry, since it does not lead to relations between $S$-matrix elements. Gauging a symmetry does not give any additional conserved charges beyond  those  of the corresponding global symmetry.
Minimal coupling is not a symmetry of any kind.

\subsection{Minimal Coupling in the SM}

The Lagrangian of the SM is a renormalizable Lagrangian.  The only terms with dimension $\le 4$ containing derivatives in the ungauged theory are the kinetic energy terms of the matter fields, and they give the correct gauged terms
\begin{align}\label{5.1}
\overline \psi\, i \slashed{D}\, \psi, \qquad D_\mu \phi^\dagger D^\mu \phi, 
\end{align}
on using minimal coupling. In this example,  minimal coupling gives the correct result, but the success is \emph{accidental}, because there are no fermion or scalar gauge interactions  of dimension $\le 4$ involving field strengths.  The situation is similar to 
$(B-L)$ conservation in the SM, which is an accidental symmetry. The most general gauge invariant Lagrangian with terms of dimension $\le 4$ automatically preserves $(B-L)$. However, if one includes higher dimension operators, the symmetry can be violated.
Similarly, in QED, there can be dimension-five anomalous magnetic moment interactions which are not given by minimal coupling. 

Eq.~(\ref{5.1}) often is given as an example of the success of minimal coupling, but in fact, renormalizable interactions in the Standard Model violate the minimal coupling principle. The reason is that there are two other interactions, the gauge kinetic term $G_{\mu \nu}^a G^{a\mu \nu}$, and a topological term proportional to $G_{\mu \nu}^a \widetilde G^{a\mu \nu}$, which both violate minimal coupling.\footnote{In pure Yang-Mills theory, the Lagrangian is only these two terms, which is a challenge for minimal coupling.}
Thus, minimal coupling actually fails even for a renormalizable gauge theory such as the SM, if one looks at the full Lagrangian.
The kinetic energy term has coefficient $-1/4$, and is order unity. The topological term has coefficient $ \theta g^2/(32\pi^2)$, and is normalized so that the action is $\theta \nu$, where $\nu$ is an integer. The topological action is not loop suppressed, even though
it has a $1/(16\pi^2)$ in the Lagrangian, because for classical field configurations (instantons), it is an integer.

It might be argued that minimal coupling
is of some value in constructing the QED Lagrangian as it does not generate an anomalous magnetic moment (Pauli) term.
However, such claims about the use of minimal coupling 
in gauge theories have long been appreciated to be incorrect~\cite{weinberglectures}.  Weinberg in his 1970 Brandeis lectures shows that for a charged fermion $\Psi$,
\bea
\frac{1}{\Lambda} \, \bar{\Psi} \, \sigma_{\mu \, \nu} \left[ \partial^\mu \,,\, \partial^\nu \right] \, \Psi \rightarrow \frac{iqe}{\Lambda} \, \bar{\Psi} \, \sigma_{\mu \, \nu} F^{\mu \,\nu} \, \Psi.
\eea
under (next-to-)minimal coupling, the same argument used in Sec.~\ref{sec:minimal}.
Minimal coupling alone does not forbid the appearance of  non-renormalizable operators involving the field strength, or lead to the conclusion that such operators must be suppressed.  Instead, as Weinberg argues,    
it is \emph{renormalizability} which forbids these terms in the SM.  
From the EFT point of view, if we want our theory to be valid to some high scale $\Lambda \to \infty$, then the Pauli operator must be suppressed by an inverse power of the scale $\Lambda$ by the EFT power counting expansion.  

\subsection{Charged Particles in an Electromagnetic Field}

Now, let us examine how minimal coupling can fail as an organizing principle.
We start with the quantum mechanical Hamiltonian for a non-relativistic particle,
\begin{align}
\mathcal{H} &= \frac{\mathbf{p}^2}{2m} \mathbf{1},
\label{5.3}
\end{align}
where $\mathbf{1}$ is a unit matrix of dimension $(2s+1)$ for a particle of spin $s$. The Hamiltonian for the interaction with the electromagnetic field is
\begin{align}
\mathcal{H}_{\text{em}} &= \frac{ \bm{\pi}^2 }{2m} \mathbf{1}=  \frac{ \left(\mathbf{p}-q e \mathbf{A}\right)^2}{2m} \mathbf{1} + e q A^0 \mathbf{1} 
\label{5.4}
\end{align}
using the substitution of Eq.~(\ref{min}), assuming the particle has charge $q$. The $e q A^0$ term arises from using Eq.~(\ref{min2}) for the $i\partial/\partial t$ term in the time-dependent Schr\"odinger equation.  Eq.~(\ref{5.4}) describes a gauge invariant theory, but is it the correct theory? For electrons (with $q_e=-1$), it is not, because it is missing the magnetic moment term.  Let us rewrite Eq.~(\ref{5.3}) for spin-1/2 particles in the alternate form
\begin{align}
\mathcal{H} &= \frac{\left(\bm{\sigma} \cdot \mathbf{p}\right)^2}{2m} =\frac{\mathbf{p}^2}{2m} \mathbf{1}.
\label{5.5}
\end{align}
We emphasize that Hamiltonian Eq.~(\ref{5.5}) is \emph{identical} to Eq.~(\ref{5.3}).  However, applying the principle of minimal coupling to Eq.~(\ref{5.5}) gives
\begin{align}
\mathcal{H}_{\text{em}} &= \frac{\left(\bm{\sigma} \cdot \bm{\pi}\right)^2}{2m} + e q A^0 \mathbf{1} = \frac{ \left(\mathbf{p}-q e \mathbf{A}\right)^2}{2m} \mathbf{1} + e q A^0 \mathbf{1} 
- \frac{e q}{2m} \bm{\sigma} \cdot \mathbf{B},
\label{5.6}
\end{align}
not Eq.~(\ref{5.4}).
Starting with the same Hamiltonian as before, and applying exactly the same principle of minimal coupling, we have generated the Hamiltonian 
Eq.~(\ref{5.6}) with a magnetic moment interaction rather than the Hamiltonian Eq.~(\ref{5.4}) with no magnetic moment interaction!  However, neither of these Hamiltonians is the correct electron interaction Hamiltonian, which is
\begin{align}
\mathcal{H}_{\text{em}} &= \frac{ \left(\mathbf{p}-q e \mathbf{A}\right)^2}{2m} \mathbf{1} + e q A^0 \mathbf{1} 
- c \frac{ qe }{2m} \bm{\sigma} \cdot \mathbf{B},
\label{5.7}
\end{align}
where $c=1.00116  =1+\alpha/(2\pi)+\ldots$ is the electron magnetic moment.

Once again, generating an interacting theory from a non-interacting one by using minimal coupling is ambiguous.
The ambiguity arises from terms involving the commutator of two derivatives,
\begin{align}
\left[\partial_\mu,\partial_\nu\right]\psi &=0,
\label{5.8}
\end{align}
which vanishes, but which leads to the non-vanishing contribution,
\begin{align}
\left[D_\mu,D_\nu\right]\psi &= i e q_\psi F_{\mu\nu}\psi \not= 0,
\end{align}
when partial derivatives are promoted to covariant derivatives via the minimal coupling procedure. This inconsistency is the reason why minimal coupling is ill-defined in theories with a derivative expansion, such as effective field theories.

\subsection{Protons and Neutrons}

The minimal coupling enthusiast might think that Eq.~(\ref{5.7}) gives the correct value of $c$ up to a missing ``loop correction.''  Maybe all one 
needs to do is to use $\bm{\sigma} \cdot \mathbf{p}$ for spin-1/2 particles (for unspecified reasons connected with the UV theory), and one will correctly reproduce all electromagnetic interactions up to ``loop factors.''  However, this reasoning is also not correct. Consider another well-known spin-1/2 particle, the proton. Using $q_p=+1$ leads to either Eq.~(\ref{5.4}) with $c=0$ or Eq.~(\ref{5.6}) with $c=1$ depending on whether Eq.~(\ref{5.3}) or Eq.~(\ref{5.5}) is used as the starting point. The proton magnetic moment is $c=2.793$, so the correction $\delta c$ to the minimal coupling Hamiltonian is $\delta c=2.793$ or $\delta c=1.793$ depending on the starting choice. The missing term is no longer a small correction of ``loop level.'' 

Maybe all the proton and electron examples show is that one should be using next-to-minimal coupling, not minimal coupling.
However, consider another familiar spin-1/2 particle, the neutron, with $q_n=0$.  For the neutron, the ordinary and covariant derivatives are the same, and like the neutral scalar example in Sec.~\ref{sec:minimal}, one cannot generate a magnetic interaction via minimal coupling.
Nevertheless, the neutron has a magnetic moment which is approximately $-2/3$ as big as the proton,  $\mu_n=-1.91$ nuclear magnetons.  In this case, minimal coupling does not produce \emph{any} electromagnetic interactions, since $q=0$, but the neutron clearly interacts electromagnetically, and its magnetic moment interaction is not  suppressed by a loop factor.  One can prove in QCD that the ratio of the proton and neutron magnetic moments is~\cite{Dashen:1993jt,Dashen:1994qi}
\begin{align}
\frac{\mu_n}{\mu_p} &= -\frac{2}{3} + \mathcal{O}\left( \frac{1}{N_c^2} \right),
\label{npratio}
\end{align}
which follows from an exact $SU(6)_c$ symmetry in the $N_c \to \infty$ limit~\cite{Dashen:1993as,Dashen:1993jt,Jenkins:1998wy}. This is a symmetry relation that violates minimal coupling.
 The ``minimal coupling'' prediction is zero, since electromagnetic interactions are proportional to the particle charges.

\subsection{The Hydrogen Atom and Quarkonia}

Another example in which neutral particles can couple to photons with couplings which are not loop suppressed is the Hydrogen atom. 
The interaction of the neutral Hydrogen atom with an external electric field leads to an effective interaction
\begin{align}
\mathcal{H} &= - \frac12 \alpha_E \mathcal{E}^2,
\label{6.11}
\end{align}
where $\mathcal{E}$ is the electric field,  $\alpha_E=9 a_0^3/2$ is the electric polarizability,  and $a_0$ is the Bohr radius. Clearly, there is no loop suppression at work here, since there are no loop graphs in non-relativistic quantum mechanics. Brambilla, Pineda, Soto and Vairo pioneered an effective theory for QCD bound states called pNRQCD~\cite{Brambilla:1999xf,Brambilla:2004jw}, and QED bound states are a special case of their formalism. The interaction Eq.~(\ref{6.11}) with $\alpha_E=9 a_0^3/2$ is consistent with their power counting. 

Similarly, electromagnetic transitions of Hydrogen, such as the $2p \to 1s+\gamma$ transition, are not loop suppressed.  The transition amplitudes are given by $\mathbf{p} \cdot \mathbf{E}$ dipole transitions proportional to the field strength, and obey the pNRQCD power counting. Neutrality of the bound states implies that the electric fields at long distances are dipole fields of order $1/r^3$ from a 
$\mathbf{p \cdot E}$ interaction, rather than Coulomb fields of order $1/r^2$ from a $q A^0$ interaction.  Neutrality of bound states does not imply that electromagnetic transitions are suppressed by loop factors.

These basic examples are not  exceptions to a general rule. Consider another example with a non-Abelian gauge symmetry.  The interaction of quarkonium with background chromoelectric and chromomagnetic fields is given by an interaction~\cite{Luke:1992tm}
\begin{align}
\, \hspace{1cm}\mathcal{L} &= \left(P_v^\dagger P_v - V^\dagger_{\mu,v} V^\mu_v\right)\left(c_E O_E + c_B O_B\right), \nn
O_E &= - G^{A \mu \alpha} G^{A \nu}{}_\alpha v_\mu v_\nu, &
\hspace{-3cm} O_B =  \frac12 G^{A \alpha \beta}G^A_{\alpha \beta} - G^{A \mu \alpha} G^{A \nu}{}_\alpha v_\mu v_\nu, 
\end{align}
where $P_v$ annihilates the pseudoscalar meson $\eta_{c}$ or $\eta_b$, and $V_{\mu,v}$ annihilates the vector meson $J/\psi$ or $\Upsilon$ for the $Q=c$ and $b$ quarkonia systems, respectively.  Here, $v$ is the velocity of the state, and $a_0$ is its radius. In the rest frame, 
$v^\mu=(1,0,0,0)$, and 
\begin{align}
O_E &= \mathbf{E}^a \cdot \mathbf{E}^a, & O_B &= \mathbf{B}^a \cdot \mathbf{B}^a,
\end{align}
are the squares of the color electric and magnetic fields. The coefficient $c_E$, the analogue of $\alpha_E$ for the Hydrogen atom, was computed by Peskin~\cite{Peskin:1979va},
\begin{align}
c_E &= \frac{14\pi}{27} a_0^3,
\end{align}
where $a_0$ is the bound state radius. The quarkonia states $\eta_{c,b}$, $J/\psi$ and $\Upsilon$ are all color-neutral particles.  Nevertheless, they couple to colored fields with couplings given by 
pNRQCD power counting, with no loop suppression factors, contrary to the expectation from minimal coupling.

The intuitive idea that the photon must couple with a loop suppression to neutral fields, which is based on the accidental success of minimal coupling for the matter part of the Standard Model  Lagrangian, cannot be used in arbitrary EFTs.

\subsection{Baryon Chiral Perturbation Theory}

Now consider the interactions of baryons with photons at low momentum. These interactions are described by the baryon chiral Lagrangian. To describe the interactions of baryons with photons, we can study an even simpler version of the theory where the pion fields have been set to zero. The low-momentum interactions of baryons with the electromagnetic field can be described by a relativistic Lagrangian of the form
\begin{align}
\mathcal{L} &= \overline \psi \left(i \slashed{\partial}- e q \slashed{A}-m\right)\psi - \frac{e \widetilde c}{4m} \overline \psi \sigma^{\mu \nu} \psi F_{\mu \nu},
\label{5.11}
\end{align}
where the total magnetic moment is $q+\widetilde c$, with $q$ coming from the fermion kinetic term.   The baryon interactions also can be described by using a non-relativistic Lagrangian written in terms of velocity-dependent fields~\cite{Jenkins:1990jv},
\begin{align}
\mathcal{L} &= \overline \psi_v  v \cdot \left(i{\partial}- e q {A}\right)\psi_v - \frac{e  c}{4m} \overline \psi_v \sigma^{\mu \nu}
\psi_v  F_{\mu \nu},
\label{5.12}
\end{align}
where $v^\mu=(1,0,0,0)$ in the baryon rest frame and the magnetic moment is $c$.  The kinetic term in Eq.~(\ref{5.12}) does not generate a magnetic moment interaction because it is spin independent. Eq.~(\ref{5.12}) is obtained from Eq.~(\ref{5.11}) by taking the non-relativistic limit. The relativistic Lagrangian Eq.~(\ref{5.11}) with $A^\mu=0$ matches onto the non-relativistic Lagrangian Eq.~(\ref{5.12}) with $A^\mu=0$,
\begin{align}
\overline \psi \left(i \slashed{\partial}-m\right)\psi  &\to  \overline \psi_v  v \cdot \left(i{\partial}\right) \psi_v. 
\label{5.13}
\end{align}
The relativistic magnetic moment interaction matches onto the non-relativistic one,
\begin{align}
 - \frac{e  }{4m} \overline \psi \sigma^{\mu \nu} \psi F_{\mu \nu} & \to - \frac{e  }{4m} \overline \psi_v \sigma^{\mu \nu}
\psi_v  F_{\mu \nu}\,.
\end{align}
However, for the two Lagrangians to describe the same physics, we need $c=\widetilde c+q$, i.e.\ the matching of the gauged kinetic term is
\begin{align}
\overline \psi \left(i \slashed{\partial}- e q \slashed{A}-m\right)\psi  &\to \overline \psi_v  v \cdot \left(i{\partial}- e q {A}\right)\psi_v - \frac{e  q}{4m} \overline \psi_v \sigma^{\mu \nu}
\psi_v  F_{\mu \nu}\,.
\label{5.15}
\end{align}
This result (using the principle of minimal coupling and then matching) is \emph{not} the same as first matching using Eq.~(\ref{5.13}) and then using the principle of minimal coupling, which misses the $F_{\mu \nu}$ term in Eq.~(\ref{5.15}).  Introducing gauge interactions by the principle of minimal coupling is incorrect in a non-renormalizable EFT, even in a simple one such as Eq.~(\ref{5.12}), which is obtained from Eq.~(\ref{5.11}) by integrating out antiparticle states at tree level. 
Note that in our example, we have worked only at tree level, and the missing $F_{\mu \nu}$ term in Eq.~(\ref{5.15}) is \emph{not} suppressed by a loop factor.

\section{EFTs of Goldstone Bosons and Dynamical Symmetry Breaking}\label{sec:chpt}

An interesting idea for solving the hierarchy problem is to have the weak interaction gauge symmetry spontaneously broken by a strongly interacting sector, in analogy to the spontaneous breaking of (global) chiral symmetry in QCD due to strong interaction dynamics. The earliest implementations of this idea were technicolor theories, which were essentially QCD-like theories with $f$, the analogue of the pion decay constant, equal to $v$, the electroweak symmetry breaking scale. 

The $SU(3)_L \times SU(3)_R$ chiral symmetry of QCD with three light flavors is spontaneously broken by the strong interaction dynamics. As a result, one has an octet of Goldstone bosons, the $\pi$, $K$ and $\eta$ mesons. The $SU(3)_L \times SU(3)_R$ symmetry is realized non-linearly; under an infinitesimal chiral transformation $\epsilon$, the octet Goldstone boson field $\pi$ shifts by 
\begin{align}
\pi \to \pi + f \epsilon\,,
\label{pion}
\end{align}
where $\pi$ and $\epsilon$ are traceless $3 \times 3 $ matrices. The chiral symmetry Eq.~(\ref{pion}) forbids a pion mass term $\text{Tr}\, \pi \pi$, so the pions are exactly massless in the absence of interactions which explicitly break the chiral symmetry, such as quark masses and electromagnetism. 

A similar symmetry argument is applicable in composite Higgs theories when the Higgs field is generated dynamically as a pseudo-Goldstone boson. In QCD, one can make the $K$ meson doublet arbitrarily light compared to $\lqcd$ by taking $m_s \to 0$. A similar mechanism is commonly used in composite Higgs theories to generate a light scalar with the quantum numbers of the SM Higgs boson; one simply makes the symmetry breaking parameter small.  In these theories, one needs to explicitly break the Higgs shift symmetry Eq.~(\ref{pion}), otherwise Yukawa couplings to SM fermions are forbidden. 
We will directly address the issues of minimal coupling in composite Higgs theories in Subsection \ref{higgscase}. 
However, before doing this, it is instructive to study minimal coupling in $\chi$PT, which describes the interactions of the photon with neutral particles in a spontaneously broken strongly coupled gauge theory, i.e. QCD.

\subsection{Chiral Perturbation Theory}\label{sec:chiral}

The chiral Lagrangian is an effective field theory with an expansion in powers of momentum $p/\lchi$, where $\lchi \sim 1$~GeV is the scale of chiral symmetry breaking. As long as the momentum $p \ll \lchi$, one has a well-defined weakly coupled expansion.  As $p$ approaches $\lchi$, the momentum expansion breaks down, and the theory becomes strongly coupled. The lowest order Lagrangian of order $p^2$ is
\begin{align}
{\cal L}_2 &= \frac{f^2}{4} \vev{ D_\mu U D^\mu U^\dagger + U^\dagger \chi + \chi^\dagger U} + \ldots
\label{6.16}
\end{align}
where $\vev{ \cdot }$ denotes a trace over flavor indices, and
\begin{align}
U &= e^{2 i \bm{\pi}/f}, & \pi &= \pi^a T^a, & \vev{T^a T^b} &= \frac 1 2 \delta^{ab}, &
\chi &= 2B_0 M, 
\end{align}
and 
\begin{equation}
M = \left[ \begin{array}{ccc} m_u & 0 & 0 \\ 0 & m_d & 0 \\ 0 & 0 & m_s \end{array}\right]
\end{equation}
is the quark mass matrix, which is order $p^2$ in the power counting. $\chi$ is the explicit breaking of chiral symmetry due to the light quark masses, and $f \sim 93$~MeV is the pion decay constant.  Amplitudes with different numbers of pions arise from the expansion of $U$, and are related algebraically, as a consequence of the chiral symmetry.

The chiral Lagrangian is a non-renormalizable theory, because $U$ is exponential in the pion fields, and generates interactions of arbitrarily high dimension when expanded.  Thus, there is an infinite series of higher order terms that must be included to obtain a consistent theory. The order $p^4$ terms are~\cite{Pich:1995bw}
\begin{align}
{\cal L}_4 &=
L_1 \,\langle D_\mu U^\dagger D^\mu U\rangle^2 \, + \,
L_2 \,\langle D_\mu U^\dagger D_\nu U\rangle\,
   \langle D^\mu U^\dagger D^\nu U\rangle \nn 
&
+~L_3 \,\langle D_\mu U^\dagger D^\mu U D_\nu U^\dagger
D^\nu U\rangle\,
+ \, L_4 \,\langle D_\mu U^\dagger D^\mu U\rangle\,
   \langle U^\dagger\chi +  \chi^\dagger U \rangle \nn  
   &
+~L_5 \,\langle D_\mu U^\dagger D^\mu U \left( U^\dagger\chi +
\chi^\dagger U
\right)\rangle\,
+ \, L_6 \,\langle U^\dagger\chi +  \chi^\dagger U \rangle^2 \nn
 &
+~L_7 \,\langle U^\dagger\chi -  \chi^\dagger U \rangle^2\,
+ \, L_8 \,\langle\chi^\dagger U \chi^\dagger U
+ U^\dagger\chi U^\dagger\chi\rangle
\nn 
 &
-~i L_9 \,\langle F_R^{\mu\nu} D_\mu U D_\nu U^\dagger +
     F_L^{\mu\nu} D_\mu U^\dagger D_\nu U\rangle\,
+ \, L_{10} \,\langle U^\dagger F_R^{\mu\nu} U F_{L\mu\nu} \rangle
\nn  &
+~H_1 \,\langle F_{R\mu\nu} F_R^{\mu\nu} +
F_{L\mu\nu} F_L^{\mu\nu}\rangle\,
+ \, H_2 \,\langle \chi^\dagger\chi\rangle \, .
\label{l4}
\end{align}
where $F_{L,R}$ are background flavor gauge fields. For electromagnetic interactions, one substitutes 
$F_L^{\mu\nu}=F_R^{\mu \nu}=e Q F^{\mu \nu}$, where $Q=\text{diag}(2/3,-1/3,-1/3)$ is the quark charge matrix.
 
The size of coefficients in the chiral Lagrangian is given by naive dimensional analysis~\cite{Manohar:1983md}, with terms of order
\begin{align}
\sum_n f^2  \lchi^2 \ \frac{\mathcal{L}_n}{\lchi^n}  =f^2  \lchi^2 \left[ \frac{\mathcal{L}_2}{\lchi^2} +  \frac{\mathcal{L}_4}{\lchi^4} + \ldots \right]\,
= \frac{\lchi^4}{16\pi^2} \left[ \frac{\mathcal{L}_2}{\lchi^2} +  \frac{\mathcal{L}_4}{\lchi^4} + \ldots \right]\,
\label{power}
\end{align}
where $\lchi \sim 4 \pi f$ is the scale of chiral symmetry breaking. (The last equality is not valid in the large $N$ limit, see Sec.~\ref{sec:largen}). There is a double expansion, in powers of $1/\lchi$ from the derivative expansion, and in powers of $1/f$ from expanding $U$. There is no sense in which terms in Eq.~(\ref{power}) can be classified in terms of ``tree'' or ``loop'' graphs in QCD. $\lchi$ and $f$ are non-perturbative parameters, as discussed in Sec.~\ref{sec:tree}.

The anomalous dimensions of the couplings $L_i$ are 
\begin{align}
\mu \frac{\rd L_i}{\rd \mu} &= \frac{\gamma_i }{16\pi^2},
\label{6.21}
\end{align}
where $\gamma_i$ are pure numbers~\cite{Pich:1995bw}, e.g.\ $\gamma_{10}=-1/4$.  The anomalous dimensions do not contain any suppression factors which maintain an operator classification scheme based on minimal coupling. 
The operators with coefficients $L_9,L_{10}$ involve the field strength $F_{\mu \nu}$, violate minimal coupling and have anomalous dimensions of order unity.
The chiral Lagrangian leads to observable scattering amplitudes, and the coefficients $L_i$ can be determined from experiment.  The best fit values of $L_i$ are given in Ref.~\cite{Pich:1995bw,Colangelo:2010et}. A quick glance shows that there is no loop suppression whatsoever of $L_{9,10}$ in this EFT of a strongly interacting sector.  In fact, $L_{9,10}$ are slightly larger than the other coefficients.

\subsubsection{Interactions of Neutral Pseudo-Goldstone Bosons}

There are some interesting features of the interaction of the neutral pseudo-scalars with photons in $\chi$PT which are relevant for PGB Higgs models.
Consider QCD with three massless quarks, and turn on electromagnetic interactions. Even in the presence of electromagnetism, there is an unbroken chiral $SU(2)_L \times SU(2)_R \times U(1)_L \times U(1)_R$ symmetry generated by 
\begin{align}
\frac12 \left[ \begin{array}{cc} 0 & 0 \\ 0 & \tau^a
\end{array} \right],\qquad
\frac{1}{\sqrt{12}} \left[ \begin{array}{ccc} -2 & 0 & 0  \\  0 & 1 & 0 \\
0 & 0 & 1
\end{array} \right].
\end{align}
This symmetry is spontaneously broken, so the $\pi^0$, $\eta$, $K^0$ and $\bar K^0$ are exact Goldstone bosons, and have a shift symmetry Eq.~(\ref{pion}).  It is important to realize that terms such as
\begin{align}
\left(\pi^0\right)^2 F_{\mu \nu} F^{\mu \nu}, \qquad K^0 \bar K^0 F_{\mu \nu} F^{\mu \nu}
\end{align}
of order $p^4$, with non-derivative interactions, are forbidden by this unbroken exact chiral symmetry, not because the mesons are neutral.  When the  remaining exact chiral symmetry is broken, e.g. for $K^0$ and $\bar K^0$ by turning on $m_s$, or for $\pi^0$ by non-zero $m_u$ and $m_d$, then one generates
\begin{align}
m_{u,d} \left(\pi^0\right)^2 F_{\mu \nu} F^{\mu \nu}, \qquad  m_s K^0 \bar K^0 F_{\mu \nu} F^{\mu \nu}.
\end{align}
interactions from the order $p^6$ terms~\cite{Bijnens:1999sh}\cite{Fearing:1994ga} in the chiral Lagrangian without any loop suppression. We have explicitly verified that this is the case. Thus, neutral pseudo-Goldstone bosons in QCD do couple to photons without a loop factor once the Goldstone boson shift symmetry is broken.

\subsection{The Higgs as a Pseudo-Goldstone Boson}\label{higgscase}

Now consider a very interesting class of models in which the Higgs is a pseudo-Goldstone boson~\cite{Weinberg:1972fn,Kaplan:1983fs,ArkaniHamed:2001nc,Georgi:2007zza}. A strongly interacting gauge theory with scale $\Lambda_S$ spontaneously breaks its
global symmetry group $G$ down to a subgroup $H$.  The Goldstone bosons are parametrized by a field $\Sigma(x)$ that lives in the coset space $G/H$ \cite{Coleman:1969sm,Callan:1969sn}.  One can weakly gauge a subgroup $G_W=SU(2) \times U(1)$ of $G$ or a larger subgroup that contains the SM $SU(2) \times U(1)$ group.  The $G$ symmetry is broken explicitly by these weak gauge couplings, and the Goldstone bosons are no longer exactly massless, but develop a potential proportional to powers of the weak gauge coupling. 
The theory is constructed so that a pseudo-Goldstone multiplet with the quantum numbers of the SM Higgs doublet develops a vacuum expectation value $v \ll \Lambda_S$, which breaks $SU(2) \times U(1)$ down to $U(1)$ electromagnetism.

In PGB Higgs models, the leading sigma model interaction terms are 
\bea
\mathcal{L}_{2} =  \frac{f^2}{2} (D^\mu \Sigma)^\dagger (D_\mu \Sigma ) + \ldots
\label{8.1}
\eea
There are also all possible higher dimension operators consistent with the EFT expansion in powers of $1/\Lambda_S$. The PGB Higgs theory, just like any other EFT, is not minimally coupled. The $p^2$ Lagrangian is only accidentally minimally coupled, because of the structure of the kinetic term in Eq.~(\ref{8.1}).  In general, the higher order terms will not be minimally coupled, and they include terms containing field-strength tensors, just as they do in QCD. The idea that PGB Higgs theories are minimally coupled arises from ignoring the $\ldots$ in Eq.~(\ref{8.1}), and the gauge kinetic terms.\footnote{It is important to point out that the original papers developing the PGB Higgs idea~\cite{Kaplan:1983fs,ArkaniHamed:2001nc,Georgi:2007zza} are not based on minimal coupling arguments. The concept was introduced in later work.}

Of particular relevance  is the neglect of operators of the form
\bea
\mathcal{O}_{W \, \Sigma} &=& ig_2 \, W_a^{\mu \, \nu}  {\rm Tr} \left[ (D_\mu \Sigma)^\dagger \,  \tau^a \, 
D_\nu \Sigma\right],  \nn
\mathcal{O}_{B \, \Sigma} &=& ig_1 \, B^{\mu \, \nu}  {\rm Tr} \left[ (D_\mu \Sigma)^\dagger \,  D_\nu \Sigma\, \tau^3\right],
\label{6.12}
\eea
which,  when expanded out, give contributions to operators  that have been argued to be loop suppressed in Ref.~\cite{Elias-Miro:2013gya}, but without any loop suppression factor.
As we have shown, one cannot invoke a minimal coupling loop factor to suppress these operators. Eq.~(\ref{6.12}) is written for theories in which the Higgs field $H$ is part of $\Sigma(x)$, and $\Sigma(x)$ transforms as a $(2,2)$ under weak and custodial $SU(2)$ symmetry. In more general $G/H$ symmetry breaking patterns, operators of the form Eq.~(\ref{6.12}) are present, but the notation is more abstract --- $\tau^a/2$ and $\tau_3$ should be thought of as the gauge generators acting on $\Sigma(x) \in G/H$.

\begin{figure}
\centering
\begin{tabular}{cc}
\includegraphics[width=4cm]{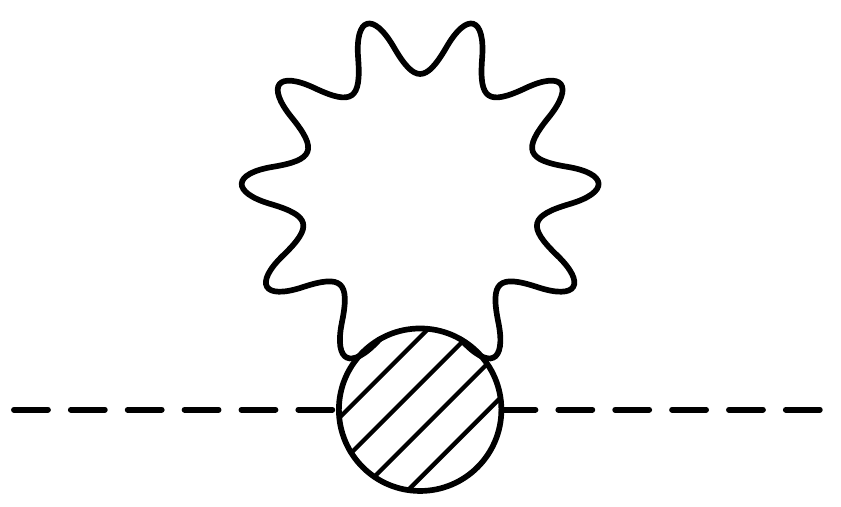} & \includegraphics[width=4cm]{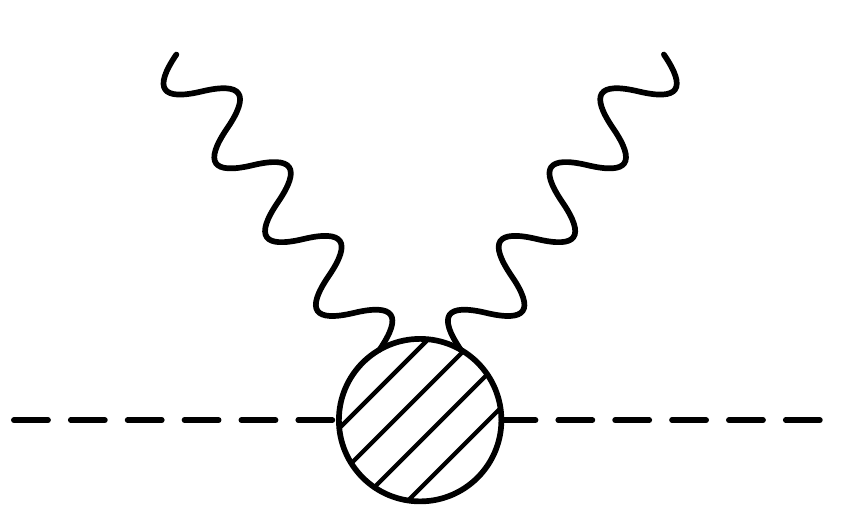} \\
(a) & (b)
\end{tabular}
\caption{\label{fig:pion} (a) Gauge boson contribution to the pseudo-Goldstone boson mass $m_H^2 H^\dagger H$ (b) 
Higgs-gauge interaction $H^\dagger H F_{\mu \nu}F^{\mu \nu}$.}
\end{figure}

In PGB models, the Higgs would be an exactly massless Goldstone boson if the SM couplings were turned off, because the Goldstone boson's shift symmetry $H \to H + v \epsilon$ is analogous to the pion chiral shift symmetry $\pi \to \pi + f \epsilon$ in QCD. In QCD, weakly gauging electromagnetism generates a mass for the $\pi^+$ through graphs shown in Fig.~\ref{fig:pion}(a), which generates the
Goldstone boson potential
\begin{align}
\mathcal{L}_M &= c\ \text{Tr}\, U Q U^\dagger Q ,
\label{7.14}
\end{align}
where $c$ is of order $f^2 \lchi^2 \times e^2 /(16 \pi^2)$. We can use the perturbative estimate of $c$ since the EFT is weakly coupled below the scale $\lchi$, and the integral is cut off at $\lchi$. Eq.~(\ref{7.14}) gives a pion mass term of order $m^2 \sim \lchi^2 \times e^2 /(16 \pi^2)$, and is of the order of $m_{\pi^+}^2-m_{\pi^0}^2$. This term explains the $\pi^+ - \pi^0$ mass splitting in QCD.\footnote{The total $\pi^+$ mass also gets a contribute from the $u$ and $d$ quark masses.} 

The general structure of Fig.~\ref{fig:pion}(a) involves the amplitude for the PGB to interact with two gauge fields,
\begin{align}
g^2 X^{ab}\left(\Sigma, \ldots\right)
\end{align}
where $g$ is the gauge coupling, which generates the mass term from Fig.~\ref{fig:pion}(a),
\begin{align}
\mathcal{L}_M &= \frac{g^2}{16\pi^2} \Lambda_S^2 X^{aa}\left(\Sigma, \ldots\right) \to - m_H^2 H^\dagger H + \ldots
\label{7.16}
\end{align}
as for QCD, where we have used the  cutoff $\Lambda_S$.
The $\ldots$ indicates that $X^{ab}$ is an arbitrary invariant polynomial that can be expanded in powers of the symmetry breaking interactions of the theory, such as gauge interactions $gT^a$, or quark mass terms $\chi$.  The precise form for $X^{ab}$ depends on the $G \to H$ symmetry breaking pattern and the way in which $SU(2) \times U(1)$ is embedded in $G$.

In QCD, if we weakly gauge a subgroup of the $SU(3)_V$ symmetry, $X^{ab}$ has the form
\begin{align}
X^{ab} &= c_1 g^2  \text{Tr}\, U T^a U^\dagger T^b + c_2 g^4 \text{Tr}\, U  T^a U^\dagger  T^b U  T^c U^\dagger  T^c
+c_3  g^2  \text{Tr}\, \chi T^a U^\dagger T^b + \ldots\,,
\label{7.16a}
\end{align}
to show some representative terms. For the case of electromagnetic interactions, we have only one generator $T^a \to Q$ and $g \to e$.
The $c_1$ term in Eq.~(\ref{7.16a})  generates the $\pi^+$ and $K^+$ electromagnetic masses, the $c_2$ term generates higher order corrections to the $\pi^+$ and $K^+$ masses, and the $c_3$ term generates a mass for the neutral $K^0$ meson of order $e^2 m_s$.
In Little-Higgs theories, the Higgs mass is of order $g^4$, so $\ldots$ in the $X^{ab}$ contains an additional factor of $(g T)(g T)$, and there is an additional gauge interaction hidden in the blob in Fig.~\ref{fig:pion}.
The details of $X^{ab}$ do not matter. By assumption in some PGH models, an interaction Eq.~(\ref{7.16}) generates the Higgs mass. But this implies (by cutting open the gauge loop in Fig.~\ref{fig:pion}(a) to get Fig.~\ref{fig:pion}(b)) that the EFT also has a Higgs-gauge interaction allowed by the symmetries of the theory,
\begin{align}
\mathcal{L} &=  \frac{g^2 c_6}{\Lambda_S^2} H^\dagger H F_{\mu \nu} F^{\mu \nu}\ .
\label{7.18}
\end{align}
Closing the gauge loop in Fig.~\ref{fig:pion}(b) to generate Eq.~(\ref{7.16}) shows that
\begin{align}
m_H^2 &\sim \frac{g^2\Lambda_S^2}{16\pi^2}c_6\,.
\label{7.19}
\end{align}
The relationship discussed here is between operators of the form in Eq.~(\ref{7.18}),  and the resulting counter terms generated by the divergent loops, which generate operators of the form that contribute to the Higgs mass.
In Little Higgs theories, $c_6$ would be order $g^2$, and $m_H^2$ order $g^4$.  It does not matter what suppression factors $m_H$ and $c_6$ have; we know experimentally that $m_H \sim 125$~GeV.
From Eq.~(\ref{7.19}), we see that (approximately)
\begin{align}
c_6 \sim \frac{16 \pi^2 m_H^2}{g^2 \Lambda_S^2} \sim \left( \frac{2.5\, \hbox{TeV}}{\Lambda_S}\right)^2
\end{align}
and $c_6$ is not  suppressed by a loop factor for $\Lambda_S$ in the $1-10$~TeV range. This equation also provides a rough relation between $\Lambda_S$ and any observed deviation of the $h \to \gamma\gamma$ rate from the SM value due to the operator in Eq.~(\ref{7.18}).

\subsection{Higher Dimension Higgs Operators}
 
In Ref.~\cite{Grojean:2013kd}, we considered the subset of dimension-six operators
\begin{equation}
\begin{aligned}
\mathcal{O}_{GG} &=  {g_3^2} \, H^\dagger \,  H \, G_{\mu\, \nu}^A G^{A\, \mu \, \nu}, & \hspace{1cm}
\widetilde{\mathcal{O}}_{GG} &=  {g_3^2} \, H^\dagger \,  H \,  G_{\mu\, \nu}^A \widetilde G^{A\, \mu \, \nu}, \\
\mathcal{O}_{BB} &=  {g_1^2} \, H^\dagger \,  H \, B_{\mu\, \nu} B^{\mu \, \nu}, & 
\widetilde{\mathcal{O}}_{BB} &=  {g_1^2} \, H^\dagger \,  H \, {B}_{\mu\, \nu} \widetilde B^{\mu \, \nu}, \\
\mathcal{O}_{WW} &=  {g_2^2} \, H^\dagger \,  H \, W^a_{\mu\, \nu} W^{a\,\mu \, \nu}, &
\widetilde{\mathcal{O}}_{WW} &=  {g_2^2} \, H^\dagger \,  H \, {W}^a_{\mu\, \nu}  \widetilde W^{a\,\mu \, \nu}, \\
\mathcal{O}_{WB} &=  {g_1 \, g_2} \, H^\dagger \, \tau^a \, H \, W^a_{\mu \, \nu} B^{\mu\, \nu},  &
\widetilde{\mathcal{O}}_{WB} &=  {g_1 \, g_2} \, H^\dagger \, \tau^a \, H \, W^a_{\mu \, \nu} \widetilde B^{\mu\, \nu}  ,
\label{ops}
\end{aligned}
\end{equation}
of the 59 dimension-six operators in the operator basis of Ref.~\cite{Grzadkowski:2010es}.
The operators in Eq.~(\ref{ops}) were chosen because they contribute to $gg \to h$, $h \to \gamma \gamma$, and $h \to Z \gamma$ at tree-level in the EFT, i.e. through tree diagrams in the EFT which are linear in the dimension-six operator coefficients. 
The unknown coefficients $c_i/\Lambda^2$ accompanying the operators parametrize arbitrary new physics.  For this formalism to be model independent, no assumption should be made about the size of the operator coefficients beyond the usual assumption of the EFT power counting.   A simple model that yields these operators is given in the appendix.

In the operator basis of Ref.~\cite{Grzadkowski:2010es}, 15 of the 59 dimension-six operators do not involve fermions.  These 15 operators are the eight operators of Eq.~(\ref{ops}),
three additional operators involving the Higgs doublet field,
\begin{align}
&\mathcal{O}_\varphi =\left(H^\dagger H\right)^3 ,
&&\mathcal{O}_{\varphi \square} =\left(H^\dagger H\right) \partial^2 \left(H^\dagger H \right),
&&\mathcal{O}_{\varphi D}=\left(H^\dagger D_\mu H\right)^\dagger \left( H^\dagger D_\mu H\right)  \,.
\label{1.7}
\end{align}
and four additional pure-gauge operators
\begin{align}
\mathcal{O}_{GGG} &= f^{ABC} G^A_{\mu}{}^\nu G^B_{\nu}{}^\rho G^C_{\rho}{}^\mu,  
&\mathcal{O}_{GG \tilde G}&=  f^{ABC} \widetilde G^A_{\mu}{}^\nu G^B_{\nu}{}^\rho G^C_{\rho}{}^\mu, \nn  
\mathcal{O}_{WWW} &= \epsilon^{abc}  W^a_{\mu}{}^\nu W^b_{\nu}{}^\rho W^c_{\rho}{}^\mu,   
&\mathcal{O}_{WW\widetilde W}&=\epsilon^{abc} \widetilde W^a_{\mu}{}^\nu W^b_{\nu}{}^\rho W^c_{\rho}{}^\mu .
\end{align}
A full calculation of the $15 \times 15$ anomalous dimension matrix needs to be performed.  However, it is easy to construct models of new physics which only produce the operators $\mathcal{O}_{WW}$, $\mathcal{O}_{BB}$ and 
$\mathcal{O}_{WB}$ in the EFT at the matching scale $\Lambda$.  For such models, the RG running of the operator subset yields  the results for $h \rightarrow \gamma \gamma$ and $h \rightarrow Z \gamma$ given in our paper~\cite{Grojean:2013kd}.

The authors of Ref.~\cite{Elias-Miro:2013gya} argue that the choice of operator basis is extremely important.  They consider the operator 
$\mathcal{O}_{BB}$
and four additional operators 
\begin{equation}
\begin{aligned}
\mathcal{P}_{HW} &=  -i \, g_2 \, (D^\mu H)^\dagger \, \tau_a \, (D^\nu H) \,  W^a_{\mu \, \nu}, & \hspace{1cm}
\mathcal{P}_{HB} &=  -i \, g_1 \, (D^\mu H)^\dagger \,  (D^\nu H) \,  B_{\mu \, \nu},\\
\mathcal{P}_{W} &=  -\frac{i \, g_2}{2} \, (H^\dagger \,  \tau_a \, \overleftrightarrow{D}^\mu H) \,  (D^\nu W^a_{\mu \, \nu}), & \hspace{1cm}
\mathcal{P}_{B} &=  -\frac{i \, g_1}{2} \, (H^\dagger \, \overleftrightarrow{D}^\mu H) \,  (D^\nu B_{\mu \, \nu}), 
\end{aligned}
\label{ops2}
\end{equation}
none of which are in the basis of Ref.~\cite{Grzadkowski:2010es}.
The $\mathcal{P}$ operators are related by the equations of motion to $\mathcal{O}_{BB}$, $\mathcal{O}_{WW}$ and $\mathcal{O}_{WB}$ by
\begin{align}
\mathcal{P}_B &= \mathcal{P}_{HB} + \frac14 \mathcal{O}_{BB}+\frac14 \mathcal{O}_{WB}, \quad \quad \mathcal{P}_W = \mathcal{P}_{HW} + \frac14 \mathcal{O}_{WW}+\frac14 \mathcal{O}_{WB}.
\label{7.5}
\end{align}
Notice that two of the three operators of Ref.~\cite{Grojean:2013kd}, namely $\mathcal{O}_{WW}$ and 
$\mathcal{O}_{WB}$, have been eliminated from this subset of five operators. 
Ref.~\cite{Elias-Miro:2013gya} uses the basis of Ref.~\cite{Giudice:2007fh}.  This basis is viewed to be a special transparent basis in  
which operators obey a tree and loop suppression organization in the EFT when minimal coupling is invoked in the underlying theory.  In this basis, one combination of the operators $\mathcal{P}_{HW}$ and $\mathcal{P}_{HB}$ gives a direct coupling of the photon to $ZH$ in the EFT (a violation of minimal coupling), and another combination of the operators contributes to the gyromagnetic ratio of the $W$ deviating from $2$, which is believed to not happen at ``tree'' level in the EFT due to minimal coupling. If another basis was used where more operators contributed to these effects, there would be no coherent scheme to use to identify offending operators directly, hence the belief that this basis is important. This reasoning underlies the opinion expressed in Ref.~\cite{Elias-Miro:2013gya} that for broad scenarios satisfying a minimal coupling assumption, the large contribution to the Wilson coefficients of the operators $\mathcal{P}_{HW}$,  $\mathcal{P}_{HB}$, and
$\mathcal{O}_{WB}$ in the EFT must cancel in mixing that gives $h \rightarrow \gamma \gamma$.  It is asserted that any direct contribution to these operators is loop suppressed, due to minimal coupling in the underlying theory, and that the only large contribution that these operators can possibly receive comes from $\mathcal{P}_B$ and $\mathcal{P}_W$, which they claim do not mix with the single Wilson coefficient $\mathcal{O}_{BB}$ giving $h \rightarrow \gamma  \gamma$ in their basis.

Ref.~\cite{Elias-Miro:2013gya} considers operator mixing of their five operators but does not calculate explicitly the anomalous dimension matrix for this entire subset of operators.  Instead, they use the anomalous dimension calculation    
of Ref.~\cite{Grojean:2013kd}, and they argue
that some of the remaining unknown anomalous dimensions vanish because ``tree'' operators do not mix with ``loop'' operators.  
These zeros and the results of Ref.~\cite{Grojean:2013kd}
are used to determine part of the  $5 \times 5$ anomalous dimension matrix for the alternative set of 5 operators, 
$\mathcal{O}_{WW}$, $\mathcal{O}_{BB}$, $\mathcal{O}_{WB}$, 
$\mathcal{P}_{HW}$ and $\mathcal{P}_{HB}$, which they call the GJMT basis of Ref.~\cite{Grojean:2013kd}, despite the fact that it involves  $\mathcal{P}$ operators, and was not the basis used in Ref.~\cite{Grojean:2013kd} for the anomalous dimension calculation.

Two of the new operators introduced in Ref.~\cite{Elias-Miro:2013gya}, $\mathcal{P}_{W}$ and $\mathcal{P}_{B}$, can be eliminated using the gauge field equations of motion, $D^\nu W_{\mu \nu}^a=g_2 j_\mu^a$ and $D^\nu B_{\mu \nu}=g_1 j_\mu$, in terms of the gauge currents which involve fermion and Higgs fields. Operators with a similar structure, the penguin operators
$\overline\psi \gamma^\mu T^A P_L \psi\, \left( D_\nu G_{\mu \nu}^A \right)$, arise in the study of non-leptonic weak interaction, and are eliminated in favor of four-quark operators using the equation of motion $D^\nu G_{\mu \nu}^A=g_3 j_\mu^A$ in standard treatments of hadronic weak decays.

We have explained at length in this paper why minimal coupling operator classification schemes are invalid in EFT.
Ref.~\cite{Elias-Miro:2013gya} uses such a scheme to argue that $h \rightarrow \gamma \gamma$ is loop-suppressed in the EFT, and that this loop suppression is preserved under RG running.

\section{Other Misconceptions}\label{sec:other}

In this section, we discuss the invalidity of other arguments which have been advanced in the literature, and are related to the minimal coupling arguments discussed earlier.

\subsection{Confining Large $N$ Theories are not like the Linear Sigma Model}\label{sec:largen}

Strongly interacting gauge theories in the large $N$ limit produce a weakly coupled spectrum of mesons. Three-meson interactions are order $1/\sqrt N$, four-meson interactions are $1/N$, etc. In theories such as large $N$ QCD with a spontaneously broken flavor symmetry, one can construct a sigma model description as in chiral perturbation theory.  The $N \to \infty$ limit of the gauge theory is taken so that the analog of 
$\lqcd$, or equivalently, meson masses are order unity as $N \to \infty$. We will call this generic scale $m$.

\begin{figure}
\centering
\includegraphics[angle=90,width=3cm]{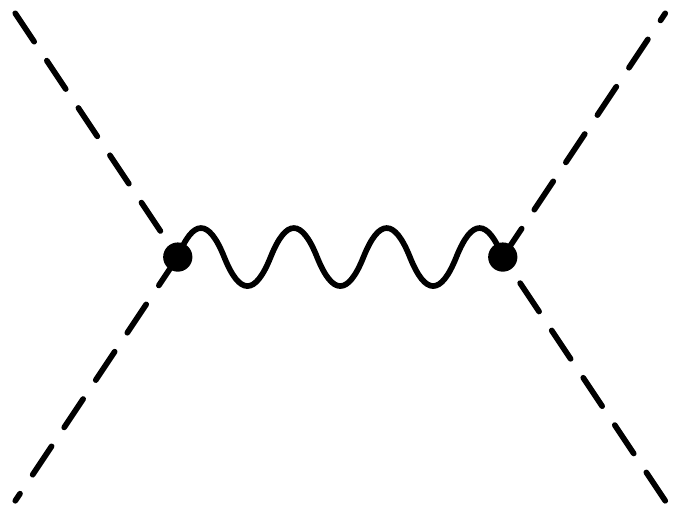}
\caption{\label{fig:pipi} Meson exchange contribution to $\pi-\pi$ scattering.}
\end{figure}

The Goldstone boson dynamics is described by a chiral Lagrangian that depends on $f \propto \sqrt N$, so that $f \sim \sqrt{N} m$. Since the Lagrangian is expanded in powers of $\pi/f$, it automatically implements the suppression by $1/\sqrt N$ for each additional meson. The Lagrangian has an expansion in derivatives suppressed by $\lchi$, which is order unity, not order $\sqrt N$, in the large $N$ limit. The scale of the derivative expansion is given by the scale of variation of form factors, or by the masses of mesons, which are order unity. 
One way to see this is to consider Fig.~\ref{fig:pipi}, which is a contribution to $\pi-\pi$ scattering due to the exchange of a meson.
The amplitude is order $1/N$, from the two $1/\sqrt N$ factors at each three-meson vertex. The expansion of $1/(p^2-m^2)$, produces terms of the form $\left(p^2/m^2\right)^r$, so there is a $p^4/(N m^4)$ contribution to the scattering amplitude. This is precisely the amplitude from the $p^4$ term
\begin{align}
\frac{f^2}{\lchi^2} \vev{\partial_\mu U \partial^\mu U^\dagger \partial_\nu U \partial^\nu U^\dagger} \sim 
\frac{f^2}{\lchi^2} \left( \frac{p}{f} \right)^4 \pi^4 + \ldots \sim \frac{p^4}{N m^4} \pi^4 + \ldots
\end{align}

The large $N$ theory also has an infinite tower of mesons with arbitrarily high spin. The reason is that correlation functions of gauge invariant operators can be obtained by summing over physical intermediate states, by the optical theorem. The gauge theory correlation functions at momentum $Q$ scale with powers of $\log Q$, by asymptotic freedom. The only way this $\log Q$ behavior can be reproduced by the sum over mesons is if the sum is over an infinite number of terms. One also must have mesons of arbitrarily high spin, since one can construct gauge invariant operators of arbitrarily high spin in the fundamental theory, such as $\overline \psi\gamma^{\mu_0} D^{\mu_1} \ldots D^{\mu_n} \psi$, and these have to couple to physical mesons so that the optical theorem is satisfied. 

The 't~Hooft model, large $N$ QCD in $1+1$ dimension, has been exactly solved, and the above points have been explicitly demonstrated. The meson form factors can be computed, and vary on a scale of order one~\cite{Einhorn:1976uz}. Operator correlation functions are reproduced by a sum over intermediate mesons~\cite{Grinstein:1997xk}, a feature often referred to as quark-hadron duality. A linear sigma model with a fundamental scalar where the scalar self-coupling $\lambda$ is large does not capture all of the important physics of a confining gauge theory, which has an infinite tower of states.

\subsection{$g\not=2$}\label{highenergy}

One argument which has been advanced for the loop suppression of higher dimensional operators violating minimal coupling is that
the value $g=2$ for the gyromagnetic ratio of fundamental fields arises at tree level, and operators with explicit factors of $F_{\mu \nu}$ can lead to tree level deviations from this value.  We have already seen how $g=2$ arises in a fundamental renormalizable 
theory~\cite{weinberglectures,Ferrara:1992yc} --- the anomalous magnetic moment operators are suppressed by $1/\Lambda$ and vanish as 
$\Lambda \to \infty$.  Arguments of this form do not imply that there are no $1/\Lambda$ terms for finite $\Lambda$.  Several examples already have been discussed where these terms occur without loop suppression.

A second argument advanced to argue that some higher dimension operators ($\mathcal{P}_{HW}$ and $\mathcal{P}_{HB}$) in the EFT are loop suppressed is that they can contribute to scattering amplitudes of photons with $W$ and $Z$ bosons, potentially leading to high-energy behavior of scattering amplitudes which violates unitarity. In an EFT, one only needs to ensure that unitarity is not violated below the scale $\Lambda$ of new physics. 
General arguments~\cite{Georgi:1985kw} imply that $\mathcal{P}_{HW}$ and $\mathcal{P}_{HB}$ do not lead to violations of unitarity in the EFT until the scale $\Lambda$.  This result is true even for massive gauge bosons such as the $W$ and $Z$, provided the gauge symmetry is spontaneously broken.  We have verified by explicit calculation that the cross section $\gamma+W \to \gamma+W$ in the EFT does not violate unitarity below $\Lambda$, even if $\mathcal{P}_{HW}$ and $\mathcal{P}_{HB}$ have coefficients which are not suppressed by $g^2/(16\pi^2)$.

\subsection{The UV Theory Cannot be Adjusted to Enforce Minimal Coupling in the IR}

It has been argued that the underlying UV theory can be chosen so that the EFT is minimally coupled. This is usually done using the mantra ``we assume that the UV completion is such that the EFT is minimally coupled.''
Consider the implications of this in QCD, as an example of a prototypical UV strongly interacting theory. 
We will take the light quark masses $m_{u,d,s}$ to be zero, as they are irrelevant to the discussion.  The fundamental theory has only two parameters, the scale $\lqcd$ and electromagnetic coupling constant $e$.  In the low-energy EFT, the $\rho$ mass determines $\lqcd$, and the $\rho^+$ charge determines $e$.  There are no other parameters in the full QCD theory which can be adjusted to enforce minimal coupling in the low-energy effective theory.  Or another example --- let the low energy theory be atomic physics with the UV theory given by QED.  What does it mean to adjust the UV theory so that atomic interactions satisfies minimal coupling, and Hydrogen polarizabilities and transition amplitudes are loop suppressed?  There is no such freedom to modify QED.  It certainly cannot be done by any local modification.

To make the argument even more  explicit, consider QED bound states in a potential $V(\mathbf{x})$. We know that $V(\mathbf{x})$ must be a Coulomb potential, but assume that our ``UV theory'' is sufficiently flexible that we can adjust  $V(\mathbf{x})$ arbitrarily, while still retaining Coulomb bound states. Can we adjust $V(\mathbf{x})$ to make Hydrogen and Positronium minimally coupled? For this to happen, all electromagnetic dipole transition matrix elements must be ``loop suppressed.'' The oscillator-strength sum-rule for the dipole transition operators is~\cite{Cohen-Tannoudji:1977fk}
\begin{align}
1 &= \sum_f \frac{2 m \left(E_f-E_i\right)}{2 \hbar^2} \abs{\braket{f | x^a | i}}^2, \qquad a=1,2,3\,.
\end{align}
One cannot make  the $i \to f$ dipole transition matrix elements highly suppressed for all $f$, regardless of the potential $V(\mathbf{x})$.

The concept of choosing the UV theory so that the low-energy effective theory is minimally coupled is generally invalid; a UV theory will contain \emph{fewer} parameters than the EFT, which has an infinite set of gauge invariant operators if one works to arbitrarily high orders in 
$1/\Lambda$.  There are not enough adjustable parameters in the UV theory to make the low energy theory  minimally coupled. One can assume the EFT has symmetries such as $CP$ which follow from symmetries of the underlying UV theory because they constrain low-energy scattering amplitudes through Ward identites. \emph{However, minimal coupling is not a  symmetry, and does not give such constraints.}

\section{Conclusions}\label{sec:concl}

We have examined the concept of minimal coupling in EFT, and shown that it is contradicted by experimental evidence and explicit calculations in a wide range of examples. The results should not be surprising to those who have applied EFT methods to analyze experimental measurements of strong, weak and electromagnetic processes. Minimal coupling is not a symmetry, and does not lead to constraints on the low-energy effective theory other than the usual one of gauge invariance.  In general, the use of minimal coupling as an organizing principle in an EFT is simply invalid and inconsistent.  The notion that a UV theory can be chosen so that the EFT is minimally coupled also is misguided.

We have shown that in the case of PGB Higgs models, 
many claims in the literature based on minimal coupling  are not  generic predictions.  The impact of new physics on the properties of the SM Higgs can be described by an EFT which contains all higher dimensional operators consistent with the symmetries of the theory with operator coefficients suppressed
by the scale of new physics $\Lambda$.  There is, in general, no additional ordering due to minimal coupling of the operator coefficients beyond the usual power counting expansion and gauge invariance of the EFT.

\acknowledgments
This research was supported in part by the Department of Energy through DOE Grant No. 
DE-FG02-90ER40546. We thank B. Gavela, G. Isidori and W. Skiba for comments on the manuscript.

\appendix

\section{Generating ``Loop-Level'' Operators at Tree Level}

In this paper, we have shown that ``non-minimally coupled'' operators in EFTs are not loop suppressed in general, and we have shown in Sec.~\ref{higgscase} how operators such as $\mathcal{P}_{HW}$,  $\mathcal{P}_{HB}$, and the $h \to \gamma \gamma$ amplitude, can arise at tree-level in PGB Higgs models.

Here we note another mechanism by which ``loop'' operators can be generated with large coefficients consistent with the EFT power counting expansion.  
A strongly interacting theory can generate a low-energy EFT description that has an infinite tower of states with arbitrary spin, 
and a tower of interactions consistent with the EFT's power counting, including super-renormalizable operators with explicit factors of the $\Lambda$ scale of the EFT. Integrating out low mass states in the EFT can generate the desired operators with order unity coefficients.

A simple toy example is given by the interaction Lagrangian
\begin{align}
\mathcal{L} &= \frac{1}{\Lambda} \left(a g_2^2 \sigma W^{a\, \mu \,\nu} \, W^a_{\mu \, \nu} + b g_1^2 \sigma B^{\mu \,\nu} B_{\mu \,\nu} + c g_1g_2\, \Sigma^a W_{\mu \,\nu}^a B^{\mu \,\nu}\right) + \Lambda \left(d \, H^\dagger H \sigma +  \, f \, H^\dagger \tau^a H \Sigma_a \right),
\end{align}
where $\sigma$ and $\Sigma^a$ are real scalar fields with masses $m_\sigma$ and $m_\Sigma$, respectively, and $W^a_{\mu\nu}$ and $B_{\mu\nu}$
are the usual ${SU}(2)$ and $U(1)$ field strengths of the SM. This example is an EFT with a cutoff scale
$\Lambda$.  The particles $\sigma$ and $\Sigma^a$ are generated by dynamics at the scale $\Lambda$. An exactly solvable model which gives this toy model is constructed in Ref.~\cite{Manohar:2013rga}.

Integrating out $\sigma$ and $\Sigma_a$ at tree level gives the effective interaction Lagrangian
\begin{align}
\mathcal{L} &= \frac{ a d}{m_\sigma^2} g_2^2 H^\dagger H \, W^{a\, \mu \,\nu} \, W^a_{\mu \, \nu} + \frac{b d}{m_\sigma^2} g_1^2 H^\dagger H \, B^{\mu \,\nu} B_{\mu \,\nu} +  \frac{c f}{m_\Sigma^2} g_1g_2\, H^\dagger \tau^a H \, W_{\mu \,\nu}^a B^{\mu \,\nu} \ .
\end{align}
Since $m_\sigma,m_\Sigma \sim \Lambda$, this toy example gives the $\mathcal{O}_{WW}$, $\mathcal{O}_{BB}$ and $\mathcal{O}_{WB}$  terms expected by the EFT power counting with order one coefficients.

This toy example illustrates two points that we wish to comment on.
First, it shows that the gauge invariant operators $\mathcal{O}_{WW}$, $\mathcal{O}_{BB}$ and $\mathcal{O}_{WB}$  considered in 
Ref.~\cite{Grojean:2013kd} can receive independent contributions  consistent with EFT power counting.
Second, is the possibility of  super-renormalizable operators  being accompanied by positive powers of the EFT scale. 
When such operators are included in the EFT, one can generate any dimension-six Higgs operator at tree-level  in  a similar manner as the toy example.

\bibliographystyle{JHEP}
\bibliography{Higgs}

\end{document}